\documentclass{emulateapj} 

\usepackage{apjfonts}
\usepackage{amsmath}
\usepackage{natbib}
\usepackage[english]{babel}
\usepackage{color}
\usepackage{hyperref}
\usepackage{float}

\begin{document}
%\received{}
%\accepted{}
%\revised{}
%\slugcomment{ApJ, accepted (\today)}
\newcommand{\LeeZinn}{\mathcal{L}}
\newcommand{\msait}{MmSAIt}

\shortauthors{A.~A.~R. Valcarce, M. Catelan, J. Alonso-Garc\'ia, C. Cort\'es, J.~R. De Medeiros} 
\shorttitle{Constraints on Helium Enhancement in M4}

\title{Constraints on Helium Enhancement in the Globular Cluster M4 (NGC~6121): \\ The Horizontal Branch Test}

\author{A.~A.~R. Valcarce\altaffilmark{1,2}}
\author{M. Catelan\altaffilmark{2,3,4}}
\author{J. Alonso-Garc\'ia\altaffilmark{3,4}}
\author{C. Cort\'es\altaffilmark{5}}
\author{J.~R. De Medeiros\altaffilmark{1}}

\altaffiltext{1}{Universidade Federal do Rio Grande do Norte, Departamento de F\'isica, 59072-970 Natal, RN, Brazil}
\altaffiltext{2}{Pontificia Universidad Cat\'olica de Chile, Centro de Astroingenier\'ia, Av. Vicu\~na Mackena 4860, 782-0436 Macul, Santiago, Chile}
\altaffiltext{3}{Pontificia Universidad Cat\'olica de Chile, Instituto de Astrof\'isica, Facultad de F\'isica, Av. Vicu\~na Mackena 4860, 782-0436 Macul, Santiago, Chile}
\altaffiltext{4}{The Milky Way Millennium Nucleus, Av. Vicu\~{n}a Mackenna 4860, 782-0436, Macul, Santiago, Chile}
\altaffiltext{5}{Universidad Metropolitana de Ciencias de la Educaci\'on, Facultad de Ciencias B\'asicas, Departamento de F\'isica, Av. Jos\'e Pedro Alessandri 774, Santiago, Chile}

\date{Received April 23, 2013; accepted December 4, 2013} 

\begin{abstract}
Recent pieces of evidence have revealed that most, and possibly all, globular star clusters are composed of groups of stars that formed in multiple episodes with different chemical compositions. In this sense, it has also been argued that variations in the initial helium abundance ($Y$) from one population to the next are also the rule, rather than the exception. In the case of the metal-intermediate globular cluster M4 (NGC~6121), recent high-resolution spectroscopic observations of blue horizontal branch (HB) stars (i.e., HB stars hotter than the RR Lyrae instability strip) suggest that a large fraction of blue HB stars are second-generation stars formed with high helium abundances. In this paper, we test this scenario by using recent photometric and spectroscopic data together with theoretical evolutionary computations for different $Y$ values. Comparing the photometric data with the theoretically-derived color-magnitude diagrams, we find that the bulk of the blue HB stars in M4 have $\Delta Y \lesssim 0.01$ with respect to the cluster's red HB stars (i.e., HB stars cooler than the RR Lyrae strip)~-- a result which is corroborated by comparison with spectroscopically derived gravities and temperatures, which also favor little He enhancement. However, the possible existence of a minority population on the blue HB of the cluster with a significant He enhancement level is also discussed.
\end{abstract}

%
%________________________________________________________________

\section{Introduction}
\label{intro}

In recent years, several studies have revealed the existence of multiple stellar populations with different chemical compositions in all globular clusters (GCs) studied to date \citep[see][for a recent review and extensive references]{Gratton_etal2012}. In this context, the O-Na anticorrelation is one of the most persistent features observed in GCs, and has been (along with the Mg-Al anticorrelation) associated to different star formation episodes \citep[e.g.,][]{Carretta_etal2009}. Broadly speaking, it is believed that the first generation of stars of a GC is typically O-rich and Na-poor, but after massive \citep{Decressin_etal2007} and/or intermediate-mass stars \citep{DAntona_etal1983} eject processed material that is mixed with pristine gas, a second generation of stars is formed with an O-poor and Na-rich chemical composition \citep{Renzini2008, Decressin_etal2007b, DErcole_etal2008, Carretta_etal2010, Conroy_Spergel2011, Valcarce_Catelan2011}. This is supported by the evidence that the O-Na anticorrelation is present down to the turn-off point and main-sequence loci \citep{Gratton_etal2001}, thus ruling out the possibility that these chemical patterns may be ascribed to internal mixing in the individual stars themselves.

With the discovery of the multiple main sequences in $\omega$~Centauri \citep[NGC~5139;][]{Bedin_etal2004, Bellini_etal2009, Bellini_etal2010} and NGC~2808 \citep{Norris2004, DAntona_etal2005, Piotto_etal2005, Piotto_etal2007}, the inclusion of the initial helium abundance ($Y$) as a free parameter in low-mass stellar evolutionary calculations has again become popular, since variations in the helium abundance, due to its well-known impact upon main-sequence stars \citep[e.g.,][]{Demarque1967, Iben_Faulkner1968, Simoda_Iben1968, Simoda_Iben1970}, can produce the observed split main sequence loci \citep[e.g.,][]{Norris2004,Piotto_etal2005}. From a nuclear astrophysics perspective, such He abundance variations appear rather appealing, since the O-Na and Mg-Al variations occur naturally in the process of proton-capture nucleosynthesis, even though~-- importantly~-- the actual level of accompanying He enhancement remains unclear at present \citep[see][for a recent review and references]{Catelan2013}.

Apart from the main sequence, He abundance variations also lead to changes that can be observed along the whole color-magnitude diagram (CMD) of GCs \citep[e.g.,][]{DAntona_etal2002,Valcarce_etal2012}. One of the most important effects induced by an increase in $Y$ is observed at the horizontal branch (HB), where~-- for a given increase in $Y$~-- the luminosities of red and blue HB stars (hereafter RHB and BHB stars, respectively) increase, while the luminosities of the extreme blue HB (eBHB) stars decrease. However, the level of He enrichment is not expected to be uniform all along the HB. Due to the fact that helium-rich stars evolve more rapidly than their low-$Y$ counterparts, and thus have lower masses for a given GC age, it is expected that He-enriched stars will be located at a bluer HB location than He-``normal'' ones, on average. With this in mind, chemically anomalous stars (e.g., O-poor, Na-rich, and possibly He-rich) may be located at the BHB for metal-intermediate GCs, as is predicted by the multiple formation episodes scenario. If so, and as a consequence of the significant dependence of HB luminosity on $Y$, these BHB stars will also be significantly brighter than the (presumably O-rich, Na-poor, He-``normal'') RHB stars in the same cluster. 

Even though theoretical simulations have shown that these chemical variations in the light element (e.g., O-Na and Mg-Al) can exist in stars formed with material previously processed by the primordial generation of stars plus pristine gas, the amount of He in these new formed stars depends on several parameters as the initial mass of the donors, the amount of pristine gas, and, if this gas is accreted from outside the GC, the time when the accretion process starts, among other factors \citep[e.g.,][]{DErcole_etal2012}. In particular, and as noted by the referee, Figure~8 in \citet{DErcole_etal2012} reveals that an appropriate and dynamically reasonable combination of star formation and dilution of pristine gas with material ejected by asymptotic giant branch (AGB) and super-AGB stars allows to build up a second generation in a GC such as M4 (NGC~6121) with a small increase in helium with respect to the first generation.

Unfortunately, the initial helium abundances of most GC stars cannot be measured spectroscopically, due to the absence of photospheric helium absorption lines at temperatures lower than 8\,000~K. For this reason, only the surface helium abundances of BHB and eBHB stars can be directly measured. Stars hotter than the so-called ``Grundahl jump'' \citep{Grundahl_etal1999}, however, are dramatically affected by helium diffusion \citep[e.g.,][and references therein]{Michaud_Richer_Richard2008}, and so only HB stars in the relatively narrow temperature range $8\,000 \lesssim T_{\rm eff}[{\rm K}] \lesssim 11\,500$ can have photospheric He abundances measured that can be more or less directly traced to the star's initial helium abundance. Even in this case, however, caution must also be exercised, since~-- as is well known~-- canonical stellar models predict an increase in the surface helium abundance ($Y_s$) of stars after the so-called first dredge-up episode on the red giant branch \citep[RGB;][]{Sweigart_Gross1978}, which however may be partly compensated for by the action of helium diffusion on the main sequence \citep[][]{Proffitt_Vandenberg1991}. Non-canonical effects \citep[e.g.,][]{Sweigart_Mengel1979,DAntona_Ventura2007} may also play a role in defining the actual difference between $Y$ and $Y_s$ on the upper RGB and HB phases. 

In the particular case of M4, a metal-intermediate GC, the chemical peculiarities observed along the RGB (\citealt{Marino_etal2008} [hereafter M08], \citealt{Villanova_Geisler2011}; and references therein) and HB \citep[][hereafter M11]{Marino_etal2011} support the hypothesis of several formation episodes inside this GC. In particular, these studies associate the RHB component to the cluster's first population and the BHB component to the second one, since RHB stars are found to be O-rich and Na-poor while BHB stars are Na-rich and O-poor. What about their corresponding He abundance levels? 

While it is impossible to directly measure the photospheric He abundances of RHB stars, \citet[][hereafter V12]{Villanova_etal2012} have carried out high-resolution spectroscopy of M4 stars using UVES on ESO's VLT2, which allowed them to estimate photospheric helium abundances for 6 BHB stars with temperatures lower than 9\,500~K. Based on those measurements, and on an analysis of the positions of the stars on a CMD, the authors have concluded that M4's BHB component has a relatively large level of He enhancement, compared with RHB stars in the same cluster~-- namely, $\Delta Y \approx 0.02-0.04$. 

While the presence of significant He enrichment among M4's BHB stars may seem like a natural result, to some extent it may also be seen as somewhat surprising. In particular, as already stated, feasible scenarios do exist that provide but a minor level of He enhancement for M4 \citep{DErcole_etal2012}, and so a large $\Delta Y$ for this cluster is not a foregone conclusion. 

Comparison with the case of M3 (NGC~5272) may also offer some hints in this regard \citep{Catelan2013}. M3, like M4, also possesses well-developed RHB and BHB components. Both clusters have nearly the same masses, to within the errors \citep{mclvdm05}, and the difference in metallicity between them does not exceed 0.4~dex \citep{Carretta_etal2009b}. Yet, according to the high-resolution spectroscopic measurements carried out by \citet{Sneden_etal2004} for a large sample of RGB stars in M3, this cluster~-- unlike M4~-- contains a substantial population of ``super-oxygen-poor'' stars, i.e., stars which have undergone very extensive levels of oxygen processing, as revealed by their $[{\rm O/Fe}] < 0$. One consequence of this fact is that the interquartile range (IQR) of the [Na/O] distribution is significantly higher in M3 than it is in M4 (i.e., 0.599 vs. 0.373; \citealt{rgea10}). \citet[][their Fig.~28]{rgea10} have recently suggested a possible correlation between the level of He enhancement $\Delta Y$ and the IQR parameter, in the sense that GCs with lower IQR values should have smaller $\Delta Y$. Therefore, the measured light-element abundance patterns would favor a lower $\Delta Y$ for M4 than for M3.  

What constraints does this set on the {\em absolute} $\Delta Y$ values for M4? For M3, according to \citet[][hereafter C09]{Catelan_etal2009}, the level of He enhancement is at most very modest, namely $\Delta Y \lesssim 0.01$~-- except perhaps for the hottest $< 5\%$ among M3's HB stars.\footnote{\citet{Dalessandro_etal2013} have recently favored a higher $\Delta Y$ value among M3's BHB stars than found in C09. Their analysis is based on near-UV observations, but for the relevant temperatures the visual bandpasses of the Str\"omgren and Johnson filter systems (as used in C09 and the present study, respectively) have a much more straightforward interpretation, since they are largely immune from the (age along the HB track, $Y$) degeneracy that affects near-UV diagrams. We will expand on this in a forthcoming paper (Valcarce et al. 2013, in preparation).} Since the above arguments suggest that $\Delta Y({\rm M4}) < \Delta Y({\rm M3})$, we face the conclusion that $\Delta Y$ for M4 stars should likely be $< 0.01$ as well. This is clearly in conflict with the V12 spectroscopic measurements. 

\begin{figure}[t]
\centering
\includegraphics[width=08.0cm]{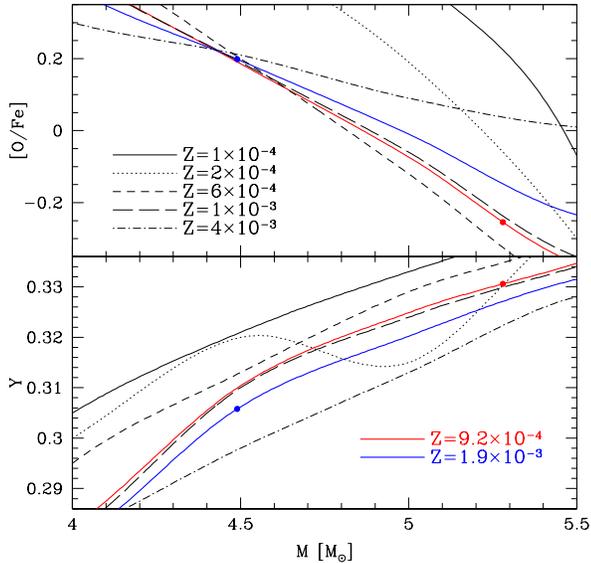}
\caption{Chemical yields of massive AGB stars from \citet{Ventura_DAntona2009} for [O/Fe] (top panel) and $Y$ (bottom panel) as a function of the initial mass. Red and blue lines represent the interpolated yields according to the metallicities of M3 (${\rm [Fe/H]}=-1.46$) and M4 (${\rm [Fe/H]}=-1.06$), respectively. Dots represent the required mass to reproduce the observed [O/Fe] values for each GC: ${\rm [O/Fe]}=0.20$ for M4 \citep{Marino_etal2008}, and ${\rm [O/Fe]}=-0.25$ for M3 \citep{Sneden_etal2004}.}
\label{FigOFeY}
\end{figure}

Figure~\ref{FigOFeY} shows the chemical yields of single-mass AGB stars (from Ventura \& D'Antona 2009), which have been suggested as potential contaminants. This figure suggests that AGB stars that contribute to a super-oxygen-poor population at the metallicity of M3 may indeed supply more He to the medium than AGB stars that contribute to an M4-like stellar population lacking super-oxygen-poor stars. While suggestive, this diagram is not conclusive by itself: obviously, it is very unlikely that single-mass AGB stars are solely responsible for the formation of a second generation of stars. Instead, in realistic chemical evolution calculations, AGB stars covering a mass range which depends on the time scale of formation of the second generation should be considered, in addition to proper dilution levels. In other words, in order to conclusively establish, in any theoretical framework, what the {\em actual} level of He enhancement in either M3 or M4 may have been, detailed models of the chemical and dynamical evolution of each cluster should be computed \citep[see, e.g.,][]{DErcole_etal2012}. In this sense, as already stated, feasible scenarios do exist that provide but a minor level of He enhancement for M4 \citep{DErcole_etal2012}. Thus, given the available evidence, it may seem difficult to accommodate the larger $\Delta Y$ level among M4's BHB stars suggested by V12. Therefore, independently verifying their spectroscopic results could be a worthwhile exercise.

The purpose of this paper is accordingly to place independent constraints on the level of He enhancement among M4's BHB stars. We use a method similar to the one used in C09 in the case of M3, where evolutionary loci computed for different $Y$ values were compared against high-precision photometric and spectroscopic observations to test for the presence of stellar populations with different helium abundances. 

This paper is structured as follows. The details of the photometric and spectroscopic data used in our work are given in Sect.~\ref{data}. In Sect.~\ref{models}, our theoretical evolutionary sequences are described. Our results are presented in Sect.~\ref{results}. Finally, our main conclusions are summarized in Sect.~\ref{conclusions}.

\vspace{0.2cm}
\section{Observational Data}
\label{data}

The photometry used in this paper is the same one as obtained in \citet{Alonso-Garcia_etal2012} for M4. This is based on observations done with the Inamori Magellan Areal Camera and Spectrograph (IMACS) in imaging mode on the 6.5m Baade Magellan telescope, and, for the central region of the cluster, on data obtained with the Advanced Camera for Surveys (ACS) camera aboard the {\em Hubble Space Telescope}. The photometry has been corrected for differential extinction with a dereddening technique that uses a non-parametric approximation to smooth the information about the reddening in the field provided by every star that belongs to the cluster \citep{Alonso-Garcia_etal2011}. This technique was successfully applied to a sample of 25 inner Galactic globular clusters in \citet{Alonso-Garcia_etal2012}.

We also use the surface gravities ($g$), effective temperatures ($T_{\rm eff}$), and chemical abundances of the HB stars in M4, as obtained by M11 and V12. 

\begin{table}
\label{TableHB}
\caption{Stellar parameters for ZAHB loci and HB evolutionary tracks at an age of 13~Gyr as a function of $Y$ and $Z$}         
\center
\begin{tabular}{@{\extracolsep{-5pt}}lcccccccc}
\hline
\multicolumn{1}{c}{$Y$}         &0.23          &0.24          &0.25          &0.26          &0.27          &0.28          &0.29          &0.32           \\  
\hline 
\hline
$Z=0.0025$                      &              &              &              &              &              &              &              &               \\  
ZAHB P. mass ($M_\odot$)& 0.865        & 0.849        & 0.834        & 0.820        & 0.805        & 0.792        & 0.778        & 0.736         \\
He core mass ($M_\odot$)        & 0.482        & 0.481        & 0.479        & 0.477        & 0.475        & 0.473        & 0.472        & 0.466         \\
HB $Y_s$                        &              & 0.255        &              &              &              & 0.292        &              & 0.330         \\
\hline
$Z=0.0017$                      &              &              &              &              &              &              &              &               \\
ZAHB P. mass ($M_\odot$)&              & 0.833        &              &              &              & 0.778        &              &               \\    
He core mass ($M_\odot$)        &              & 0.482        &              &              &              & 0.475        &              &               \\ 
\hline
\hline
\\
\\
\end{tabular}       
\end{table} 

\vspace{0.5cm}
\section{Theoretical Models}
\label{models}

The theoretical models used in this work were obtained using the Princeton-Goddard-PUC (PGPUC) code and PGPUC Online database \citep{Valcarce_etal2012}.\footnote{Available at \url{http://www2.astro.puc.cl/pgpuc/}.}  In particular, zero-age HB (ZAHB) loci were created using the PGPUC Online tool for an $\alpha$-element enhancement level given by $[\alpha/{\rm Fe}]=+0.3$, a value typical of globular cluster stars \citep{Gratton_etal2004} and close to that favored by M08 for M4, namely $[\alpha/{\rm Fe}]=+0.39 \pm 0.05$. Two heavy-elements abundances of $Z=0.0025$ and $0.0017$ were considered, in addition to seven initial helium abundances from $Y=0.23$ to $0.29$ in steps of $\Delta Y=0.01$, for a representative age of 13~Gyr \citep[e.g.,][]{Dotter_etal2010, VandenBerg_etal2013}. In addition, using the same web tool, isochrones were created for an age of 13~Gyr, both aforementioned metallicities, and $Y=0.245$. These theoretical models were transformed to the Johnson-Cousins $BVI$ passband using the bolometric corrections (BCs) and color indices from \citet{Castelli_Kurucz2003}. 

In addition to the aforementioned isochrones and ZAHB loci, full-fledged HB tracks were also computed with PGPUC, assuming $Z=0.0025$, three initial helium abundances ($Y=0.24$, $0.28$, and $0.32$), and the corresponding helium core masses (see Table~\ref{TableHB}). Based on these tracks, reference evolutionary loci were then computed as in C09, namely the middle-age HB (MAHB) and the 90\%-age HB (90AHB) loci (corresponding to the locus in the CMD occupied by model HB stars which have already lived 50\% and 90\% of their total HB lifetimes, respectively), and the terminal-age HB (TAHB) locus (corresponding to the core He exhaustion locus). One should accordingly expect to find of order 50\% of all HB stars between the ZAHB and MAHB loci, 90\% of all HB stars between the ZAHB and 90AHB loci, and 100\% between the ZAHB and the TAHB. The effects of different metallicities and initial helium abundances upon the ZAHB, 90AHB, and TAHB loci in the color-magnitude and in the theoretical $\log g$--$\log T_{\rm eff}$ diagrams were already discussed in C09.

While the present study was carried out using PGPUC models, we have checked that essentially the same results are recovered using the models from the BaSTI database \citep{Pietrinferni_etal2004}. This is not unexpected, since comparisons between the results of PGPUC and those of other widely used codes, as carried out in \citet{Valcarce_etal2012}, do not reveal large differences that might affect any of the conclusions of our paper.

\begin{figure}
\centering
\includegraphics[height=21.0cm, width=11.0cm, trim=0cm 0.3cm 8.3cm 0.5cm]{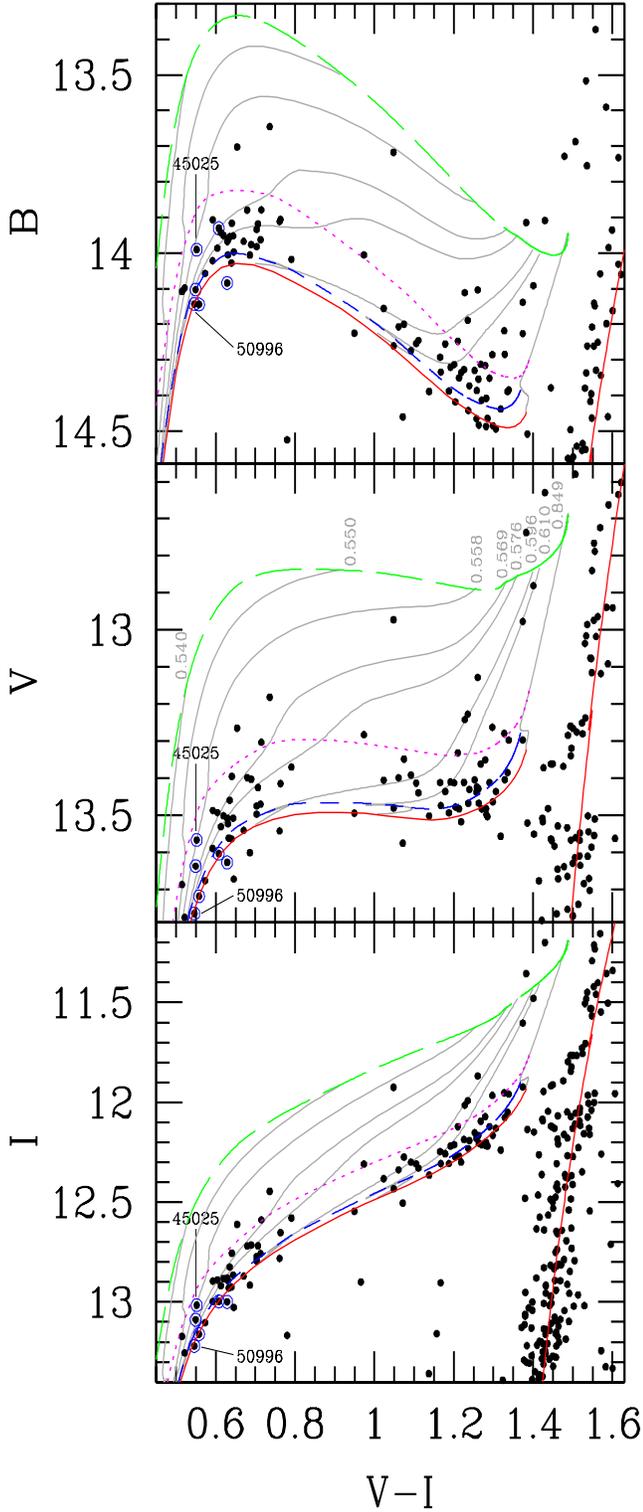}
\caption{Comparison between fiducial HB sequences and the empirical data for M4 (dots), for three different filters: $B$ ({\em upper panel}), $V$ ({\em middle panel}), and $I$ ({\em lower panel}). Lines correspond to different evolutionary HB phases for $Z=2.5\times10^{-3}$ and $Y=0.24$: ZAHB (red continuous line), MAHB (blue dashed line), 90AHB (magenta dotted line), and TAHB (green long-dashed line). Gray lines show some of the HB evolutionary tracks that were used to obtain these evolutionary loci (the corresponding mass values, in solar units, are given in the middle panel). The 13~Gyr old isochrone for the same chemical composition is also shown in each panel. Open blue circles show the HB stars studied by V12, where the two stars with lowest $\log (g)$ are indicated with their respective ID numbers. Theoretical models are corrected by reddening $E(V-I)=0.52$ mag and by a distance modulus according to Figure \ref{FigDistModulus}.}
\label{FigCMDphases}
\end{figure}

\vspace{0.5cm}
\section{ZAHB loci method in M4}
\label{results}

\begin{figure*}
\centering
\includegraphics[width=18cm, trim=0.3cm 1.0cm 0.0cm 0cm]{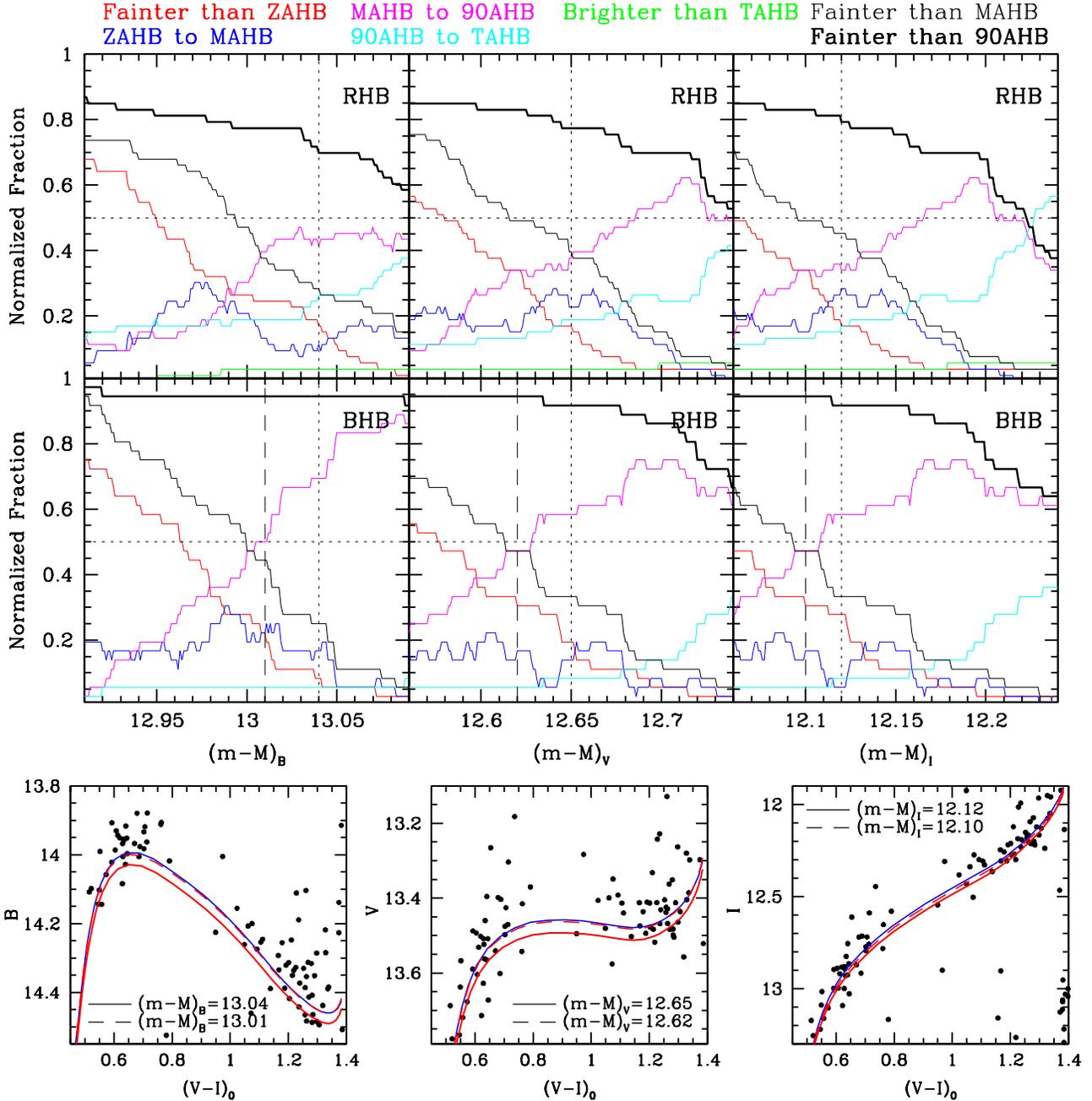}
\caption{Distance modulus determination for each filter according to the fraction of stars in each evolutionary stage along the HB, for $Z=2.5\times10^{-3}$ and $Y=0.24$ (see Figure \ref{FigCMDphases}). Upper and middle panels show the normalized fractions for stars with $(V-I)_0>0.9$ (RHB stars) and $(V-I)_0<0.9$ (BHB stars), respectively, for filters B (left panels), V (middle panels), and I (right panels). Color lines represent the normalized fraction of stars being: i)~fainter than the ZAHB locus (red lines), ii)~between the ZAHB and MAHB loci (blue lines), iii)~between the MAHB and 90AHB loci (magenta lines), iv)~between the 90AHB and TAHB loci (cyan lines), v)~brighter than the TAHB locus (green lines), vi)~fainter than the MAHB locus (thin black lines), and vii)~fainter than the 90AHB locus (bold black lines). Vertical dotted lines represent the distance modulus used in this paper for each filter selected according to the RHB population. Vertical dashed line represent the distance modulus that should be selected if only BHB stars are considered. Bottom panels: M4 CMD around the ZAHB locus. Continuous lines represent the ZAHB loci for $Y=0.24$ (red line) and $0.25$ (blue line) using the distance modulus obtained from RHB stars. Dashed red lines represent the ZAHB loci for $Y=0.24$ using the distance modulus obtained from BHB stars.}
\label{FigDistModulus}
\end{figure*}

\begin{figure*}
  \centering
\includegraphics[width=18cm, height=21cm, trim=0.5cm 0cm 0.5cm 0cm]{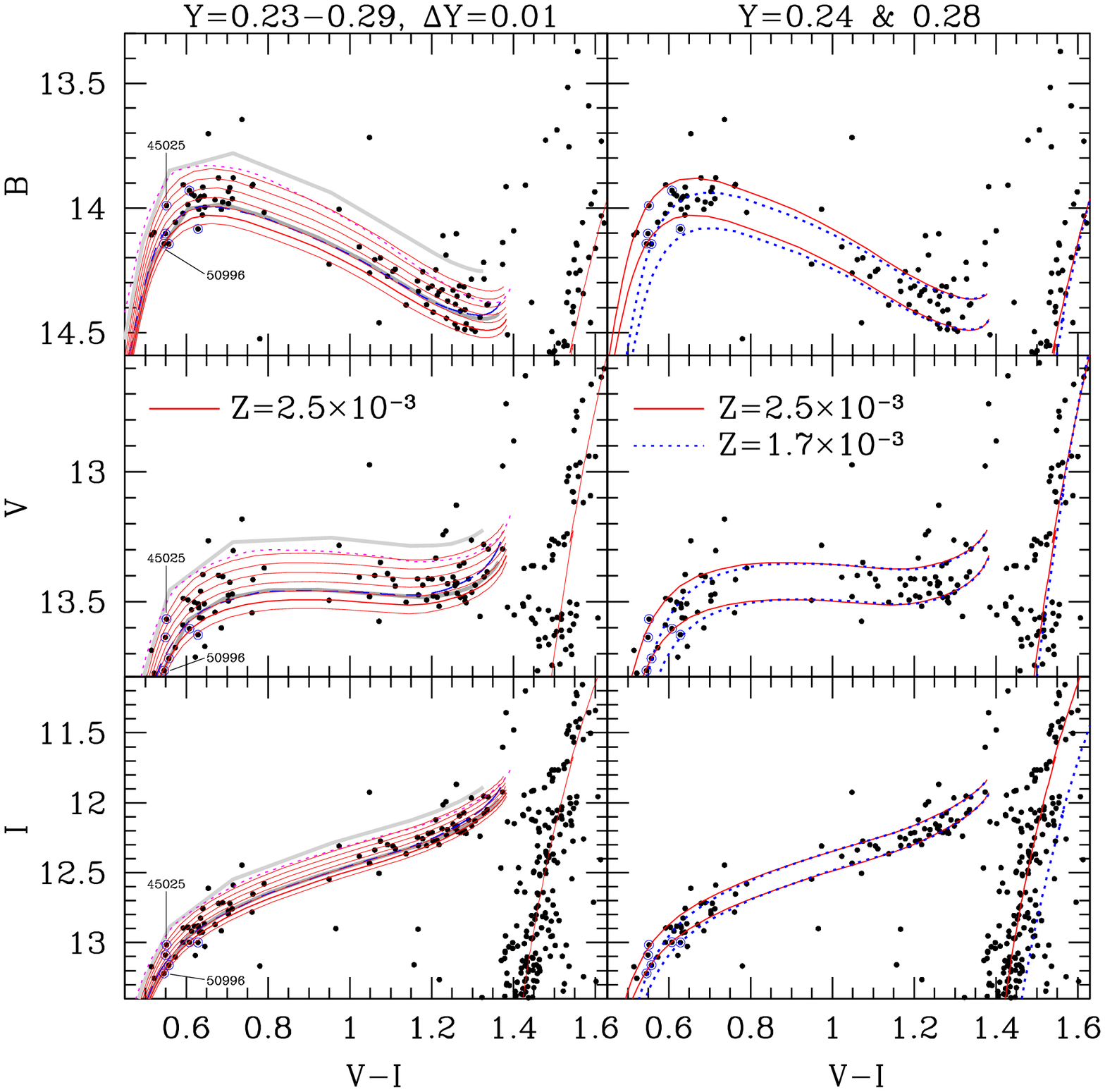}
\caption{Comparison between fiducial HB sequences and the empirical data for M4, for three different filters: $B$ ({\em upper panel}), $V$ ({\em middle panel}), and $I$ ({\em lower panel}). The {\em left panels} show the theoretical ZAHB loci for $Z=2.5\times10^{-3}$, and $Y$ from $0.23$ to $0.29$ in steps of $\Delta Y=0.01$ (continuous red lines). Dashed blue lines and dotted magenta lines show the MAHB and 90AHB loci, respectively, for $Z=2.5\times10^{-3}$ and $Y=0.24$. Gray lines show the BaSTI ZAHB loci for $Y=0.25$ and $Y=0.30$ with $Z\approx 2\times10^{-3}$. The {\em right panels} show the theoretical ZAHB loci for $Z=2.5\times10^{-3}$ (red continuous lines) and $Z=1.7\times10^{-3}$ (blue dotted line), for both $Y=0.24$ and $0.28$. In all panels is also shown the respective 13~Gyr isochrone with $Y=0.24$. Note that at the lower right panel the isochrone and the red ZAHB cannot simultaneously fit the empirical data for the lower $Z$ value. Open blue circles and numbers have the same meaning as in Fig. \ref{FigCMDphases}. Theoretical models with $Z=2.5\times10^{-3}$ are corrected as in Fig. \ref{FigCMDphases} while models with $Z=1.7\times10^{-3}$ are corrected in order to fit the red HB loci: $E(V-I)=0.56$ mag and $(m-M)_0=12.834$ mag.}
% $M_B=13.17$ mag, $M_V=12.74$ mag, and $M_I=12.20$ mag.}
\label{FigCMD}
\end{figure*}

M4 has an iron abundance with respect to the Sun of ${\rm [Fe/H]}=-1.06\pm0.01$~dex (M08, M11, V12), and no evidence of a spread in this element or overall metallicity $Z$ \citep{Carretta_etal2009b}. Therefore, we are justified in comparing observations of the cluster with models computed for a single $Z$ value. This comparison is carried out in Figure~\ref{FigCMDphases}, where we zoom in around the He-sensitive HB region. In this figure, the displayed isochrones and ZAHB/MAHB/90AHB/TAHB loci are characterized by a metallicity $Z=2.5\times 10^{-3}$, initial He abundance $Y=0.240$ (corresponding to $Y_s=0.255$ on the HB phase), and an age of 13~Gyr (corresponding to an HB progenitor mass of $0.849\, M_\odot$).\footnote{The quoted $Z$ value corresponds to the spectroscopically derived ${\rm [Fe/H]}=-1.06$, the assumed ${\rm [\alpha/Fe]}=+0.3$, and a $Y\approx 0.25$, as computed using the PGPUC Online calculator: \url{www2.astro.puc.cl/pgpuc/WebTool.Zcalculator.php}} This figure also shows some of the HB evolutionary tracks (gray lines) used for creating the ZAHB/MAHB/90AHB/TAHB evolutionary loci. As clearly shown by the figure, virtually all HB stars have luminosities between the theoretical ZAHB and 90AHB loci, in agreement with the theoretical predictions for a stellar population with a single metallicity $Z$ and He abundance $Y$. 

The distance modulus for each filter is selected according to the normalized fraction of stars at different evolutionary phases. This method is shown in Figure~\ref{FigDistModulus}, where the criteria for the choice of each distance modulus depends exclusively on the RHB stellar population. First, the normalized fraction of RHB stars is calculated in each evolutionary phase for different distance moduli (upper panels). Since photometric errors can induce variations in those fractions, we consider at the same time the normalized fraction of stars being i)~fainter than the MAHB (thin black lines), ii)~between the MAHB and the TAHB (magenta lines), and iii)~fainter than the TAHB (bold black lines). The corresponding normalized fractions should be around 50\%, 50\%, and 100\% for each of these three ranges, respectively (note that all these loci are for $Y=0.240$). 

Finally, we visually inspect the results (bottom panels in Fig.~\ref{FigDistModulus}), in order to ensure that the procedure just described produced a reasonable description of the data. As a final result, the individual estimation of the distance modulus for each filter (vertical dotted lines) gives results that are fully consistent with $(m-M)_0=12.77$~mag, as used to set the position of the ZAHB loci (red continuous lines) in the bottom panels of Figure~\ref{FigDistModulus}. In the middle panels of Fig.~\ref{FigDistModulus} the same exercise was done using only the BHB stars, giving always a distance modulus that is smaller than the one obtained with RHB stars. This can be seen in the bottom panels of this figure, where the position of the ZAHB loci with $Y=0.240$ using the distance modulus according to the normalized fraction of BHB stars (red dashed lines) is almost the same as the position of the ZAHB loci for $Y=0.250$ but using the distance modulus according to the normalized fraction of RHB stars. This suggests that M4 BHB stars are on average more He-rich compared to RHB stars, but by no more than $\Delta Y \simeq 0.01$. If indeed there is a population of helium-enriched stars occupying M4's BHB, it must however be mixed with a population of helium-normal stars, since not all BHB stars are brighter than the ZAHB locus for $Y=0.250$.

A further check for the presence of an internal variation in the He abundance is carried out in Figure~\ref{FigCMD} ({\em left panels}), where we compare ZAHB loci for different $Y$ values with the observations. In the left panels of this figure, we compare the photometric results with the theoretical ZAHB loci with $Z=2.5\times 10^{-3}$ and helium abundances from $Y=0.23$ to $0.29$, at intervals of $\Delta Y=0.01$ (continuous red lines). TAHB, 90AHB, and MAHB are not included in these plots for clarity, except for the MAHB and 90AHB loci for $Y=0.24$ (dashed blue and dotted magenta lines). As can be seen, ZAHB loci with helium abundances greater than $Y=0.27\sim0.28$ are brighter than the bulk of the blue HB stars in the cluster. Accordingly, it is highly unlikely that all blue HB stars in M4 are helium-enriched at this level. This can also be observed in the bottom panels of Figure \ref{FigDistModulus}. Of course, since the 90AHB locus is similar in brightness to the ZAHB locus for $Y=0.29$, it is possible that a few percent of the BHB stars may have an enhancement of the initial helium abundance and be confused with evolved He-``normal'' BHB stars-- but the latter do appear to constitute the bulk of the sample.

In Figure~8 of V12 the same kind of comparison as in our Figure~\ref{FigCMD} was carried out, using ZAHB loci for two initial helium abundances ($Y=0.24$ and $0.28$) and $Z= 10^{-3}$. However, their adopted $Z$ value is not consistent with the spectroscopically derived iron content of M4, since such a $Z$ value, for the same ${\rm [\alpha/Fe]} = +0.3$ adopted previously and a $Y=0.24$, corresponds to an $[{\rm Fe}/{\rm H}] = -1.47$ (or $-1.24$, if $[\alpha/{\rm Fe}] = 0.0$ is adopted instead), which is much lower than the ${\rm [Fe/H]}$ value favored by the M08, M11, and V12 spectroscopic observations, namely $-1.06 \pm 0.01$. The impact of this difference upon the inferred results can be inferred from Figure~\ref{FigCMD} ({\em right panels}), where ZAHB loci for $Y=0.24$ and $0.28$ are overplotted on the data for two different overall metallicities, namely $Z=2.5\times 10^{-3}$ and $1.7\times 10^{-3}$, corresponding to ${\rm [Fe/H]}=-1.06$ and ${\rm [Fe/H]}=-1.24$, respectively, for ${\rm [\alpha/Fe]}=+0.3$. Clearly, if a metallicity that is too low is adopted, one is led to conclude~-- erroneously, in this case~-- that the bulk of M4's HB stars are indeed He enhanced.\footnote{Note, in this sense, that the C09 statement that ``the exact choice of $Z$ value is basically irrelevant for our purposes'' meant that adoption of a metallicity slightly {\em higher} than favored by the spectroscopic data for M3 did not alter the conclusion that there seemed to be little evidence for He enhancement among M3's BHB stars.} Note, in addition, that adoption of too low a $Z$ value does not allow one to simultaneously fit the RHB and RGB loci in the $I$, $V-I$ plane ({\em bottom right} panel in Fig.~\ref{FigCMD}), as opposed to what happens when a metallicity more closely corresponding to the spectroscopic [Fe/H] value is used instead.  

\begin{figure}
\centering
\includegraphics[height=21cm, width=11.5cm, trim=0.3cm 0cm 7.5cm 0cm]{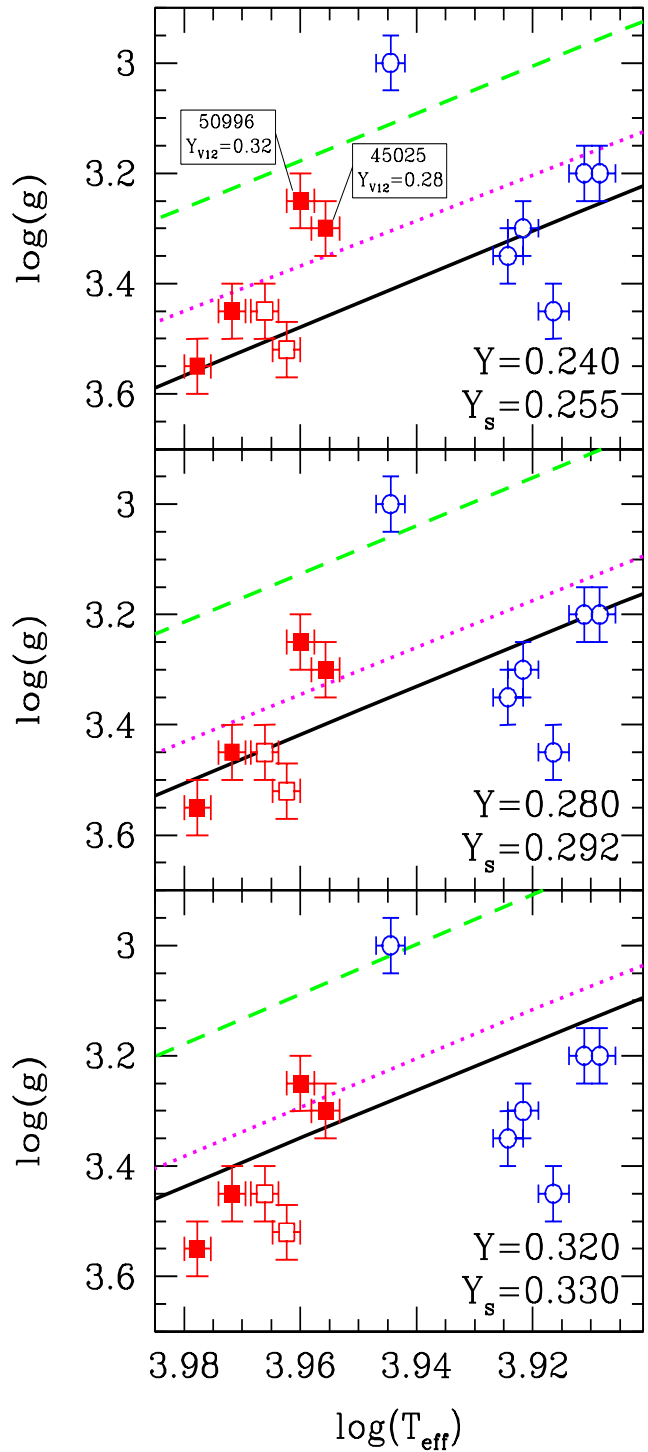}
\caption{Comparison between predicted and observed loci in the $\log g - \log T_{\rm eff}$ plane at the BHB level, for three different helium abundance values $Y$: 0.24 ({\em upper panel}), 0.28 ({\em middle panel}), and 0.32 ({\em lower panel}). In each panel is also shown the expected surface helium abundance $Y_s$. Lines represent different HB evolutionary stages: ZAHB (continuous black lines), 90AHB (dotted magenta lines), and TAHB (dashed green lines). Open and filled squares represent stars with photospheric He abundances (from V12) ${Y_{\rm V12}}=0.26$ and $0.28\le{Y_{\rm V12}}\le0.32$, respectively. In the upper panel photospheric He abundances are explicitly indicated for the 2 stars with lowest $\log g$ from V12. Circles are stars from M11 (unknown ${Y_s}$). All these stars have $[{\rm Na/Fe}]\ge0.26$ and $[{\rm O/Fe}]\le0.29$. Errors are assumed similar to the error estimated by V12, even though they could be larger (see M11).}
\label{Fig_GravTeff}
\end{figure}

As a further check of these results, and again following C09, in Figure~\ref{Fig_GravTeff} we compare the spectroscopic data and theoretical predictions for BHB stars in the $\log g - \log T_{\rm eff}$ plane. In this figure, circles and squares represent the stars studied by M11 and V12, respectively. As can be seen, 8 of the 12 BHB stars are in agreement with the theoretical predictions for a canonical helium abundance (initial $Y=0.240$ and surface helium abundance at the HB $Y_s=0.255$), 2 stars (both from M11) have empirical $g$ values outside the canonical limits by more than their stated $1\sigma$ errors, and 2 stars (both from V12) seem to be in an advanced evolutionary phase (i.e., over the 90AHB locus) for a canonical helium value. Alternatively, the latter two stars may correspond to a minority subpopulation with enhanced helium. However, only one of these stars (ID 45025) has all the properties that are asociated to stars with high initial-$Y$: high photospheric He abundance ($Y_{\rm V12}$), low surface gravity, high luminosities in the three filters studied here (B, V, and I, see Fig. \ref{FigCMD}), and high-Na and low-O abundances. As suggested by the referee, RHB stars could in principle also be included in this figure, in order to compare their surface gravities with those obtained for BHB stars and theoretical models. Unfortunately, however, these red HB stars are only studied in M11, where the $\log g$ values have an estimated error around $\pm 0.25$~dex. For this reason, it is impossible to obtain reliable information on the basis of these data alone.

Finally, it is important to note that, even though all stars in Figure~\ref{Fig_GravTeff} have Na and O abundances ($[{\rm Na/Fe}]\ge0.26$ and $[{\rm O/Fe}]\le0.29$, V12 and M11) related to stars presumably formed with material processed by an older generation of stars, almost all of them have surface gravities in agreement with canonical He abundances. This suggests that there is no direct relation between the O-Na anticorrelation and different degrees of He abundances, or~-- perhaps more plausibly~-- that the degree of He enhancement among M4's HB stars is too small to be reliably detected. This is not totally unexpected, since~-- as pointed out by \citet{Catelan2013}~-- large levels of oxygen depletion (i.e., ${\rm [O/Fe]} < 0$) may in some cases plausibly be associated with very small levels of He enhancement. In the specific case of M4, the levels of oxygen depletion detected spectroscopically are very small indeed (i.e., no stars are known with ${\rm [O/Fe]} \lesssim 0.2$), thus making it even harder to explain the presence of significant levels of He enhancement in this cluster. 

\section{Conclusions}
\label{conclusions}

In this paper, the same ZAHB loci test as previously carried out in C09 for M3 has been used to check if there are stars with high initial helium abundance in the GC M4, as recently suggested.

Using the CMD of this GC together with ZAHB loci computed for the spectroscopically derived cluster metallicity, we find that, as in the case of M3, the level of He enhancement among most BHB stars in M4 is unlikely to be higher than $\Delta Y=0.01$. This is confirmed by the distribution of the stars in the $\log g - \log T_{\rm eff}$ plane, although the latter in particular cannot rule out the presence of a minority subpopulation of stars with moderate levels of He enhancement being present in the cluster. Together with the fact that feasible theoretical scenarios do exist that provide but a minor level of He enhancement for M4 \citep{DErcole_etal2012}, a reevaluation of the V12 spectroscopic He abundance measurements would probably be worthwhile.

\vspace{0.2cm}

\acknowledgments We thank the anonymous referee for her/his comments, which have greatly helped improve the presentation of our results. Support for A.A.R.V., M.C., and J.A.-G. is provided by the Ministry for the Economy, Development, and Tourism's Programa Iniciativa Cient\'{i}fica Milenio through grant P07-021-F, awarded to The Milky Way Millennium Nucleus; by Proyecto Basal PFB-06/2007; and by Proyecto FONDECYT Regular \#1110326. A.A.R.V. and J.R.M. acknowledge additional support from CAPES grant PNPD/2011-Institucional, the CNPq Brazilian agency, and INCT INEspa\c{c}o. J.A.-G. acknowledges support from FONDECYT Postdoctoral Grant \#3130552.

\bibliographystyle{aa}

\begin{thebibliography}{51}
\expandafter\ifx\csname natexlab\endcsname\relax\def\natexlab#1{#1}\fi

\bibitem[{{Alonso-Garc{\'{\i}}a} {et~al.}(2012){Alonso-Garc{\'{\i}}a}, {Mateo},
  {Sen}, {Banerjee}, {Catelan}, {Minniti}, \& {von
  Braun}}]{Alonso-Garcia_etal2012}
{Alonso-Garc{\'{\i}}a}, J., {Mateo}, M., {Sen}, B., {et~al.} 2012, \aj, 143, 70

\bibitem[{{Alonso-Garc{\'{\i}}a} {et~al.}(2011){Alonso-Garc{\'{\i}}a}, {Mateo},
  {Sen}, {Banerjee}, \& {von Braun}}]{Alonso-Garcia_etal2011}
{Alonso-Garc{\'{\i}}a}, J., {Mateo}, M., {Sen}, B., {Banerjee}, M., \& {von
  Braun}, K. 2011, \aj, 141, 146

\bibitem[{{Bedin} {et~al.}(2004){Bedin}, {Piotto}, {Anderson}, {Cassisi},
  {King}, {Momany}, \& {Carraro}}]{Bedin_etal2004}
{Bedin}, L.~R., {Piotto}, G., {Anderson}, J., {et~al.} 2004, \apjl, 605, L125

\bibitem[{{Bellini} {et~al.}(2010){Bellini}, {Bedin}, {Piotto}, {Milone},
  {Marino}, \& {Villanova}}]{Bellini_etal2010}
{Bellini}, A., {Bedin}, L.~R., {Piotto}, G., {et~al.} 2010, \aj, 140, 631

\bibitem[{{Bellini} {et~al.}(2009){Bellini}, {Piotto}, {Bedin}, {King},
  {Anderson}, {Milone}, \& {Momany}}]{Bellini_etal2009}
{Bellini}, A., {Piotto}, G., {Bedin}, L.~R., {et~al.} 2009, \aap, 507, 1393

\bibitem[{{Carretta} {et~al.}(2009{\natexlab{a}}){Carretta}, {Bragaglia},
  {Gratton}, {D'Orazi}, \& {Lucatello}}]{Carretta_etal2009b}
{Carretta}, E., {Bragaglia}, A., {Gratton}, R., {D'Orazi}, V., \& {Lucatello},
  S. 2009{\natexlab{a}}, \aap, 508, 695

\bibitem[{{Carretta} {et~al.}(2009{\natexlab{b}}){Carretta}, {Bragaglia},
  {Gratton}, {Lucatello}, {Catanzaro}, {Leone}, {Bellazzini}, {Claudi},
  {et~al.}}]{Carretta_etal2009}
{Carretta}, E., {Bragaglia}, A., {Gratton}, R.~G., {et~al.} 2009{\natexlab{b}},
  \aap, 505, 117

\bibitem[{{Carretta} {et~al.}(2010){Carretta}, {Bragaglia}, {Gratton},
  {Recio-Blanco}, {Lucatello}, {D'Orazi}, \& {Cassisi}}]{Carretta_etal2010}
{Carretta}, E., {Bragaglia}, A., {Gratton}, R.~G., {et~al.} 2010, \aap, 516,
  A55

\bibitem[{{Castelli} \& {Kurucz}(2003)}]{Castelli_Kurucz2003}
{Castelli}, F., \& {Kurucz}, R.~L. 2003, arXiv:astro-ph/0405087

\bibitem[{{Catelan}(2013)}]{Catelan2013}
{Catelan}, M. 2013, in European Physical Journal Web of Conferences, Vol.~43,
  1001

\bibitem[{{Catelan} {et~al.}(2009){Catelan}, {Grundahl}, {Sweigart},
  {Valcarce}, \& {Cort{\'e}s}}]{Catelan_etal2009}
{Catelan}, M., {Grundahl}, F., {Sweigart}, A.~V., {Valcarce}, A.~A.~R., \&
  {Cort{\'e}s}, C. 2009, \apjl, 695, L97

\bibitem[{{Clem} {et~al.}(2004){Clem}, {VandenBerg}, {Grundahl}, \&
  {Bell}}]{Clem_etal2004}
{Clem}, J.~L., {VandenBerg}, D.~A., {Grundahl}, F., \& {Bell}, R.~A. 2004, \aj,
  127, 1227

\bibitem[{{Conroy} \& {Spergel}(2011)}]{Conroy_Spergel2011}
{Conroy}, C., \& {Spergel}, D.~N. 2011, \apj, 726, 36

\bibitem[{{Dalessandro} {et~al.}(2013){Dalessandro}, {Salaris}, {Ferraro},
  {Mucciarelli}, \& {Cassisi}}]{Dalessandro_etal2013}
{Dalessandro}, E., {Salaris}, M., {Ferraro}, F.~R., {Mucciarelli}, A., \&
  {Cassisi}, S. 2013, \mnras, 430, 459

\bibitem[{{D'Antona} {et~al.}(2005){D'Antona}, {Bellazzini}, {Caloi}, {Pecci},
  {Galleti}, \& {Rood}}]{DAntona_etal2005}
{D'Antona}, F., {Bellazzini}, M., {Caloi}, V., {et~al.} 2005, \apj, 631, 868

\bibitem[{{D'Antona} {et~al.}(2002){D'Antona}, {Caloi}, {Montalb{\'a}n},
  {Ventura}, \& {Gratton}}]{DAntona_etal2002}
{D'Antona}, F., {Caloi}, V., {Montalb{\'a}n}, J., {Ventura}, P., \& {Gratton},
  R. 2002, \aap, 395, 69

\bibitem[{{D'Antona} {et~al.}(1983){D'Antona}, {Gratton}, \&
  {Chieffi}}]{DAntona_etal1983}
{D'Antona}, F., {Gratton}, R., \& {Chieffi}, A. 1983, \memsai, 54, 173

\bibitem[{{D'Antona} \& {Ventura}(2007)}]{DAntona_Ventura2007}
{D'Antona}, F., \& {Ventura}, P. 2007, \mnras, 379, 1431

\bibitem[{{Decressin} {et~al.}(2007{\natexlab{a}}){Decressin}, {Charbonnel}, \&
  {Meynet}}]{Decressin_etal2007b}
{Decressin}, T., {Charbonnel}, C., \& {Meynet}, G. 2007{\natexlab{a}}, \aap,
  475, 859

\bibitem[{{Decressin} {et~al.}(2007{\natexlab{b}}){Decressin}, {Meynet},
  {Charbonnel}, {Prantzos}, \& {Ekstr{\"o}m}}]{Decressin_etal2007}
{Decressin}, T., {Meynet}, G., {Charbonnel}, C., {Prantzos}, N., \&
  {Ekstr{\"o}m}, S. 2007{\natexlab{b}}, \aap, 464, 1029

\bibitem[{{Demarque}(1967)}]{Demarque1967}
{Demarque}, P. 1967, \apj, 149, 117

\bibitem[{{D'Ercole} {et~al.}(2012){D'Ercole}, {D'Antona}, {Carini},
  {Vesperini}, \& {Ventura}}]{DErcole_etal2012}
{D'Ercole}, A., {D'Antona}, F., {Carini}, R., {Vesperini}, E., \& {Ventura}, P.
  2012, \mnras, 423, 1521

\bibitem[{{D'Ercole} {et~al.}(2008){D'Ercole}, {Vesperini}, {D'Antona},
  {McMillan}, \& {Recchi}}]{DErcole_etal2008}
{D'Ercole}, A., {Vesperini}, E., {D'Antona}, F., {McMillan}, S.~L.~W., \&
  {Recchi}, S. 2008, \mnras, 391, 825

\bibitem[{{Dotter} {et~al.}(2010){Dotter}, {Sarajedini}, {Anderson},
  {Aparicio}, {Bedin}, {Chaboyer}, {Majewski}, {Mar{\'{\i}}n-Franch},
  {et~al.}}]{Dotter_etal2010}
{Dotter}, A., {Sarajedini}, A., {Anderson}, J., {et~al.} 2010, \apj, 708, 698

\bibitem[{{Gratton} {et~al.}(2004){Gratton}, {Sneden}, \&
  {Carretta}}]{Gratton_etal2004}
{Gratton}, R., {Sneden}, C., \& {Carretta}, E. 2004, \araa, 42, 385

\bibitem[{{Gratton} {et~al.}(2001){Gratton}, {Bonifacio}, {Bragaglia},
  {Carretta}, {Castellani}, {Centurion}, {Chieffi}, {Claudi},
  {et~al.}}]{Gratton_etal2001}
{Gratton}, R.~G., {Bonifacio}, P., {Bragaglia}, A., {et~al.} 2001, \aap, 369,
  87

\bibitem[{{Gratton} {et~al.}(2012){Gratton}, {Carretta}, \&
  {Bragaglia}}]{Gratton_etal2012}
{Gratton}, R.~G., {Carretta}, E., \& {Bragaglia}, A. 2012, \aapr, 20, 50

\bibitem[Gratton et al.(2010)]{rgea10} 
  Gratton, R.~G., Carretta, E., Bragaglia, A., Lucatello, S., \& D'Orazi, V.\ 2010, \aap, 517, A81

\bibitem[{{Grundahl} {et~al.}(1999){Grundahl}, {Catelan}, {Landsman},
  {Stetson}, \& {Andersen}}]{Grundahl_etal1999}
{Grundahl}, F., {Catelan}, M., {Landsman}, W.~B., {Stetson}, P.~B., \&
  {Andersen}, M.~I. 1999, \apj, 524, 242

\bibitem[{{Harris}(1996)}]{Harris1996}
{Harris}, W.~E. 1996, \aj, 112, 1487

\bibitem[{{Iben} \& {Faulkner}(1968)}]{Iben_Faulkner1968}
{Iben}, I.~Jr., \& {Faulkner}, J. 1968, \apj, 153, 101

\bibitem[{{Marino} {et~al.}(2011){Marino}, {Villanova}, {Milone}, {Piotto},
  {Lind}, {Geisler}, \& {Stetson}}]{Marino_etal2011}
{Marino}, A.~F., {Villanova}, S., {Milone}, A.~P., {et~al.} 2011, \apjl, 730,
  L16

\bibitem[{{Marino} {et~al.}(2008){Marino}, {Villanova}, {Piotto}, {Milone},
  {Momany}, {Bedin}, \& {Medling}}]{Marino_etal2008}
{Marino}, A.~F., {Villanova}, S., {Piotto}, G., {et~al.} 2008, \aap, 490, 625

\bibitem[McLaughlin \& van der Marel(2005)]{mclvdm05} 
  McLaughlin, D.~E., \& van der Marel, R.~P.\ 2005, \apjs, 161, 304

\bibitem[{{Michaud} {et~al.}(2008){Michaud}, {Richer}, \&
  {Richard}}]{Michaud_Richer_Richard2008}
{Michaud}, G., {Richer}, J., \& {Richard}, O. 2008, \apj, 675, 1223

\bibitem[{{Norris}(2004)}]{Norris2004}
{Norris}, J.~E. 2004, \apjl, 612, L25

\bibitem[{{Pietrinferni} {et~al.}(2004){Pietrinferni}, {Cassisi}, {Salaris}, \&
  {Castelli}}]{Pietrinferni_etal2004}
{Pietrinferni}, A., {Cassisi}, S., {Salaris}, M., \& {Castelli}, F. 2004, \apj,
  612, 168

\bibitem[{{Piotto} {et~al.}(2007){Piotto}, {Bedin}, {Anderson}, {King},
  {Cassisi}, {Milone}, {Villanova}, {Pietrinferni}, {et~al.}}]{Piotto_etal2007}
{Piotto}, G., {Bedin}, L.~R., {Anderson}, J., {et~al.} 2007, \apjl, 661, L53

\bibitem[{{Piotto} {et~al.}(2005){Piotto}, {Villanova}, {Bedin}, {Gratton},
  {Cassisi}, {Momany}, {Recio-Blanco}, {Lucatello}, {et~al.}}]{Piotto_etal2005}
{Piotto}, G., {Villanova}, S., {Bedin}, L.~R., {et~al.} 2005, \apj, 621, 777

\bibitem[{{Proffitt} \& {VandenBerg}(1991)}]{Proffitt_Vandenberg1991}
{Proffitt}, C.~R., \& {VandenBerg}, D.~A. 1991, \apjs, 77, 473

\bibitem[{{Renzini}(2008)}]{Renzini2008}
{Renzini}, A. 2008, \mnras, 391, 354

\bibitem[{{Simoda} \& {Iben}(1968)}]{Simoda_Iben1968}
{Simoda}, M., \& {Iben}, I.~Jr., 1968, \apj, 152, 509

\bibitem[{{Simoda} \& {Iben}(1970)}]{Simoda_Iben1970}
{Simoda}, M., \& {Iben}, I.~Jr., 1970, \apjs, 22, 81

\bibitem[{{Sneden} {et~al.}(2004){Sneden}, {Kraft}, {Guhathakurta}, {Peterson},
  \& {Fulbright}}]{Sneden_etal2004}
{Sneden}, C., {Kraft}, R.~P., {Guhathakurta}, P., {Peterson}, R.~C., \&
  {Fulbright}, J.~P. 2004, \aj, 127, 2162

\bibitem[{{Sweigart}(1987)}]{Sweigart1987}
{Sweigart}, A.~V. 1987, \apjs, 65, 95

\bibitem[{{Sweigart} \& {Gross}(1978)}]{Sweigart_Gross1978}
{Sweigart}, A.~V., \& {Gross}, P.~G. 1978, \apjs, 36, 405

\bibitem[{{Sweigart} \& {Mengel}(1979)}]{Sweigart_Mengel1979}
{Sweigart}, A.~V., \& {Mengel}, J.~G. 1979, \apj, 229, 624

\bibitem[{{Valcarce} \& {Catelan}(2011)}]{Valcarce_Catelan2011}
{Valcarce}, A.~A.~R., \& {Catelan}, M. 2011, \aap, 533, 120

\bibitem[{{Valcarce} {et~al.}(2012){Valcarce}, {Catelan}, \&
  {Sweigart}}]{Valcarce_etal2012}
{Valcarce}, A.~A.~R., {Catelan}, M., \& {Sweigart}, A.~V. 2012, \aap, 547, A5

\bibitem[{{VandenBerg} {et~al.}(2013){VandenBerg}, {Brogaard}, {Leaman}, \&
  {Casagrande}}]{VandenBerg_etal2013}
{VandenBerg}, D.~A., {Brogaard}, K., {Leaman}, R., \& {Casagrande}, L. 2013, \apj, 775, 134

\bibitem[{{Ventura} \& {D'Antona}(2009)}]{Ventura_DAntona2009}
{Ventura}, P., \& {D'Antona}, F. 2009, \aap, 499, 835

\bibitem[{{Villanova} \& {Geisler}(2011)}]{Villanova_Geisler2011}
{Villanova}, S., \& {Geisler}, D. 2011, \aap, 535, A31

\bibitem[{{Villanova} {et~al.}(2012){Villanova}, {Geisler}, {Piotto}, \&
  {Gratton}}]{Villanova_etal2012}
{Villanova}, S., {Geisler}, D., {Piotto}, G., \& {Gratton}, R.~G. 2012, \apj,
  748, 62

\end{thebibliography}

\end{document}